\documentclass[aps,pre,amsmath,amssymb,notitlepage,twocolumn]{revtex4-1}

\usepackage{txfonts}
\usepackage{graphicx}
\usepackage{bm}

\begin{document}

\title{Monte Carlo simulation of classical spin models with chaotic billiards}
\author{Hideyuki Suzuki}
\email{hideyuki@mist.i.u-tokyo.ac.jp}
\affiliation{Department of Mathematical Informatics, The University of Tokyo,
             Tokyo 113--8656, Japan}
\affiliation{Institute of Industrial Science, The University of Tokyo,
             Tokyo 153--8505, Japan}

\begin{abstract}
It has recently been shown that the computing abilities of Boltzmann machines,
or Ising spin-glass models, can be implemented by chaotic billiard
dynamics without any use of random numbers.
In this paper, we further numerically investigate
the capabilities of the chaotic billiard dynamics
as a deterministic alternative to random Monte Carlo methods
by applying it to classical spin models in statistical physics.
First, we verify that the billiard dynamics can yield samples
that converge to the true distribution of the Ising model on a small lattice,
and we show that it appears to have
the same convergence rate as random Monte Carlo sampling.
Second, we apply the billiard dynamics to
finite-size scaling analysis of the critical behavior of the Ising model
and show that the phase transition point and the critical exponents
are correctly obtained.
Third, we extend the billiard dynamics to
spins that take more than two states
and show that it can be applied successfully to the Potts model.
We also discuss the possibility of extensions to
continuous-valued models such as the XY model.
\end{abstract}

\date{December 15, 2013}

\maketitle

\section{Introduction}

Many important classical spin models such as the Ising model and
the Potts model are described as probability distributions of
spin configurations.
To investigate the behavior of such models numerically,
we normally design Monte Carlo methods, and obtain samples from
the model using random numbers.
However, there is no reason in principle to use randomness
for obtaining samples;
it does not matter whether the samples are generated
randomly or deterministically,
if the samples properly represent the probabilistic models in question.

Chaotic Boltzmann machines recently proposed in Ref.~\cite{Suzuki2013b}
have chaotic billiard dynamics that yields
samples from Ising spin-glass models
without any use of random numbers.
They have been numerically shown to have computing abilities
comparable to conventional (stochastic) Boltzmann machines.
In this paper, we further numerically investigate
the capabilities of the chaotic billiard dynamics as a deterministic
alternative to random Monte Carlo methods for classical spin models.
Although there have been no studies that utilize billiard dynamics for
deterministic simulations of spin models,
the following three streams of studies can be considered as
closely related to the present study.

Firstly, several deterministic cellular automaton models
for the Ising model have been proposed.
The Q2R automaton \cite{Vichniac1984} is the simplest automaton model
of the Ising model, and it evolves conserving the energy exactly.
Creutz \cite{Creutz1986} proposed another automaton model
in which demons are introduced as additional degrees of freedom.
Demons absorb and release energy at each site conserving the total energy.
Despite the simple and deterministic update rules,
these models reproduce the probabilistic behavior of the Ising model.
However, these automata cannot be directly regarded as deterministic samplers
from the Ising model; in the Q2R automaton, spin configurations
only move on a microcanonical ensemble, and in the Creutz automaton,
the system temperature is internally determined.
Although chaotic Boltzmann machines are not automata,
there is a similarity to the Creutz automaton
in the sense that the additional degrees of freedom introduced allows
the system to deterministically generate samples from canonical ensembles.

Secondly, spin models based on coupled map lattices (CMLs) \cite{Kaneko1989}
have been proposed \cite{Sakaguchi1988,Miller1993,Marcq1997,Marcq1998,Sakaguchi1999,Egolf2000,Just2006}.
CMLs are typically composed of discrete-time chaotic elements
on a lattice interacting with each other.
They are known to exhibit rich spatio-temporal nonlinear dynamics,
and form an important class of dynamical systems
with a large number of degrees of freedom.
By associating symbols to partitions in the state space of each element,
CMLs can be regarded as deterministic Ising-like spin models.
In the sense that the probabilistic behavior of Ising spins is realized
by chaotic dynamics, CML-based spin models can be related to chaotic Boltzmann
machines. However, each element constituting chaotic Boltzmann machines
has continuous-time and non-chaotic dynamics.

Thirdly, it should be noted that random numbers
used in conventional Monte Carlo simulations
on ordinary computers, i.e., pseudo-random numbers,
are deterministically generated.
Therefore, in a sense, we have already been using
deterministic Monte Carlo methods.
Of course, such pseudo-random numbers are designed
so that they can be regarded as truly random in many aspects.
Since Monte Carlo methods rely on truly random numbers,
it is crucial to use good random numbers
in principle.
However, the generation of good random numbers is costly,
and this is one important issue in large-scale Monte Carlo simulations.
Moreover, it has been pointed out \cite{Murray2012} that
low-quality pseudo-random numbers can actually be used in Monte Carlo methods
and may even improve the performance.
Thus, it is controversial as to the quality of randomness actually required.
It is also noteworthy that recently a deterministic sampling algorithm,
which is called herded Gibbs sampling \cite{Bornn2013}, was proposed
on the basis of the herding algorithm \cite{Welling2009,Welling2010}.
Although the algorithm is not practical for large-scale simulations,
it is theoretically far more efficient than conventional Monte Carlo methods.
Thus, it is intriguing to explore what is possible
with deterministic Monte Carlo algorithms,
and chaotic Boltzmann machines can be regarded
as one of the approaches to the problem.

With these motivations in mind,
in this paper, we numerically investigate
the capabilities of the chaotic billiard dynamics
as a deterministic alternative to random Monte Carlo methods
by applying it to classical spin models in statistical physics.
We confirm that the billiard dynamics can yield samples that
converge to the true distribution of the Ising model,
and we show that the phase transition point and the critical exponents
of the Ising model are correctly obtained.
Furthermore, we extend the billiard dynamics to spins
that take more than two states,
and we apply it to the Potts model and the XY model.
We also point out that the billiard sampling dynamics is reversible
and can be a good example for discussing the microscopic origins of
irreversible macroscopic behavior on the Ising model.

It is to be noted that although the term ``Monte Carlo'' may imply that
the algorithm is probabilistic, we intentionally keep using this term
for deterministic algorithms also, thereby
indicating that they can be used exactly in the place of
conventional Monte Carlo methods.

\section{Billiard dynamics for the Ising model}

In this section, we briefly introduce the implementation of
probabilistic spin models by billiard dynamics.
Although we limit the description to the Ising model,
it can be extended to arbitrary Ising spin-glass models
or Boltzmann machines \cite{Suzuki2013b}.

Let us consider the Ising model composed of $N$ sites on a lattice.
The Hamiltonian for a spin configuration
$\bm{\sigma}=(\sigma_1,\dots,\sigma_N)\in\{-1,+1\}^N$
of the Ising model is given by
\begin{equation}
E(\bm{\sigma})=-\sum_{\langle i,j\rangle}\sigma_i\sigma_j,
\end{equation}
where the summation is taken for all the adjacent pairs in a lattice.

The probability distribution of the spin configurations
of the Ising model is given by the Gibbs distribution
\begin{equation}\label{eq:gibbs}
\mathrm{P}[\bm{\sigma}]
= \frac{1}{Z}\exp\left(-\frac{1}{T}E(\bm{\sigma})\right),
\end{equation}
where $T$ denotes the temperature and
$Z$ denotes the partition function given by
\begin{equation}
Z = \sum_{\bm{\sigma}}\exp\left(-\frac{1}{T}E(\bm{\sigma})\right).
\end{equation}

For large spin systems, it is difficult to evaluate the probability
and directly obtain the samples.
To obtain samples from the probability distribution (\ref{eq:gibbs}),
we normally use Monte Carlo methods.
Here we consider the heat-bath algorithm,
which is also known as Gibbs sampling in the field of machine learning.
We choose a spin $i$ from $1,\dots,N$ randomly or sequentially.
For each chosen spin, we update the state according to the probability
\begin{equation}\label{eq:prob}
\mathrm{P}[\sigma_i=+1\mid\bm{\sigma}_{\setminus i}]
= \frac{\exp(\sum \sigma_j/T)}{\exp(\sum \sigma_j/T)+\exp(-\sum \sigma_j/T)},
\end{equation}
where $\bm{\sigma}_{\setminus i}$ denotes the configuration of the spins
in the system except for the $i$th spin.
This process defines a Markov chain having
the Gibbs distribution (\ref{eq:gibbs}) as the stationary distribution.
Therefore, we can eventually obtain a sample sequence from
the Gibbs distribution.

Here, we consider using billiard dynamics
instead of the heat-bath algorithm
for sampling from the Gibbs distribution (\ref{eq:gibbs}).
We introduce an internal state $x_i\in[-1,+1]$ for each node $i$.
The internal state of the $i$th node evolves
according to the following differential equation:
\begin{equation}\label{eq:dx}
\frac{\mathrm{d}x_i}{\mathrm{d}t}
= -\sigma_i \exp\left(-\frac{1}{T}\sigma_i \sum_j \sigma_j\right).
\end{equation}
The state of the $i$th node $\sigma_i$ changes
when $x_i$ reaches $+1$ or $-1$ as follows:
\begin{equation}
\begin{split}\label{eq:switch}
\sigma_i \longleftarrow +1 \quad\text{when}\quad x_i= +1, \\
\sigma_i \longleftarrow -1 \quad\text{when}\quad x_i= -1.
\end{split}
\end{equation}
Note that $x_i$ decreases when $\sigma_i=+1$ and increases when $\sigma_i=-1$.
Therefore, the internal state $x_i$ continues oscillating between $+1$ and $-1$.
The continuous-time dynamics defined by
Eqs. (\ref{eq:dx}) and (\ref{eq:switch}) is
a hybrid dynamical system \cite{Aihara2010}
on the state space $\{-1,+1\}^N\times[-1,+1]^N$,
because it has both discrete and continuous state variables.

The differential equation (\ref{eq:dx}) is designed to be consistent
with Eq.~(\ref{eq:prob}) in the following sense.
We assume here that the states of the neighboring nodes of the $i$th node
are fixed.  Then, $x_i$ oscillates between $+1$ and $-1$ periodically.
According to Eq.~(\ref{eq:dx}),
it takes $2\exp(\sum \sigma_j/T)$ unit time
for $x_i$ to move from $+1$ to $-1$,
and it takes $2\exp(-\sum \sigma_j/T)$ unit time
for $x_i$ to move from $-1$ to $+1$.
Hence, the period is $2(\exp(\sum \sigma_j/T)+\exp(-\sum \sigma_j/T))$ unit time,
in which the state $\sigma_i$ takes on $+1$ for
$2\exp(\sum \sigma_j/T)$ unit time.
Therefore, the probability that we observe $\sigma_i=+1$ at a random instant is
consistent with Eq.~(\ref{eq:prob}).
Note that this consistency is derived under the assumption that
the states of the neighboring nodes are fixed.
Since the states in the system actually change,
this is merely an explanation that justifies Eq.~(\ref{eq:dx}) only intuitively.
It is expected but not assured that $\bm{\sigma}$ follows the
Gibbs distribution (\ref{eq:gibbs}).

In the explanation above,
it is only essential that the relative time duration for which
$\sigma_i$ takes on $+1$ in a period coincides with the probability
$P[\sigma_i=+1\mid\bm{\sigma}_{\setminus i}]$ defined in Eq.~(\ref{eq:prob}).
In other words, what is required for the consistency is that
the speed $\left|\mathrm{d}x_i/\mathrm{d}t\right|$ is
proportional to $P[\sigma_i\mid\bm{\sigma}_{\setminus i}]^{-1}$.
In fact, there are two other natural implementations
in place of Eq.~(\ref{eq:dx}).
One possibility is to define the speed as
$P[\sigma_i\mid\bm{\sigma}_{\setminus i}]^{-1}$ as follows:
\begin{equation}\label{eq:dxps-}
\frac{\mathrm{d}x_i}{\mathrm{d}t} =
-\sigma_i \left(1+\exp\left(-\frac{2}{T}\sigma_i \sum_j \sigma_j\right)\right).
\end{equation}
Another possibility is to define the speed as
$P[-\sigma_i\mid\bm{\sigma}_{\setminus i}]$ as follows:
\begin{equation}\label{eq:dxp-s}
\frac{\mathrm{d}x_i}{\mathrm{d}t}
= \frac{-\sigma_i \exp(-\sigma_i \sum \sigma_j/T)}
       {\exp(\sum \sigma_j/T)+\exp(-\sum \sigma_j/T)}.
\end{equation}
Considering the interactions with the neighboring nodes,
these three definitions have different dynamics.
As shown in the next section (Fig.~\ref{fig:tempsize}(a)),
all the definitions appear to work similarly,
so that we mainly employ only Eq.~(\ref{eq:dx}) in this paper.
We also note that implementations are not limited only to these three types.
For instance, arbitrary constants different from site to site
can be multiplied to the speeds.

There are two views on the dynamics of this system.
If we focus on each node in the system,
the internal state $x_i$ continues oscillating between $-1$ and $+1$,
interacting with the neighboring nodes.
In this sense, the system can be regarded as
a type of coupled oscillator system.
On the other hand, if we observe all the internal states simultaneously,
the internal state $(x_1,\dots,x_N)$ travels
in a straight line in the hypercube $[-1,+1]^N$
according to Eq.~(\ref{eq:dx}), and changes its direction when
it collides with the boundary of the hypercube
according to Eq.~(\ref{eq:switch}).
In this sense, the dynamics of this system can be regarded
as pseudo-billiard \cite{Blank2004}.

Unlike deterministic spin models implemented by CMLs,
each node does not have chaotic dynamics.
However, it has been shown \cite{Suzuki2013b,Suzuki2013a} that
chaotic behavior naturally emerges from the interactions in the system,
and it is considered to work as a heat-bath to realize probabilistic behavior
of the spin configurations

Statistics of the Ising spin model can be obtained
from the continuous-time billiard system in the following manner.
Let $t_0,t_1,\dots,t_k,\dots$ be the sequence of times at which
state switchings occur according to Eq.~(\ref{eq:switch}).
Then, the expectation value of a statistic $\Phi(\bm{\sigma})$
can be estimated from the samples until time $t_K$ as follows:
\begin{equation}
\langle\Phi\rangle
= \frac{1}{t_K - t_0} \int_{t_0}^{t_K} \Phi(\bm{\sigma}(t))\mathrm{d}t
= \frac{1}{t_K - t_0} \sum_{k=1}^{K} \tau_k \Phi(\bm{\sigma}_k),
\end{equation}
where $\tau_k=t_k-t_{k-1}$ and
$\bm{\sigma}_k$ denotes the state that the system is taking on
in the time interval from $t_{k-1}$ to $t_{k}$.
Thus, in a sense, we obtain a sample sequence
$\bm{\sigma}_1,\bm{\sigma}_2,\dots$
weighted by time intervals $\tau_1, \tau_2, \dots$.
The statistics are calculated in this manner in the present study.
Another method to obtain a (unweighted) sample sequence is to
observe $\bm{\sigma}=(\sigma_1,\dots,\sigma_N)$ uniformly,
at every one unit time for example, or randomly.
Note that sampling from a continuous-time billiard system is analogous to
sampling with continuous-time (kinetic) Monte Carlo algorithms such as
the Bortz--Kalos--Lebowitz (BKL) algorithm \cite{Bortz1975}
and the Gillespie algorithm \cite{Gillespie1977}.

Although the billiard dynamics is described as a continuous-time system,
numerical calculation can be carried out by iterating the Poincar\'e map
$\bm{x}(t_{k})\mapsto\bm{x}(t_{k+1})$
induced on the boundary of the hypercube $[0,1]^N$
as explained in Refs.~\cite{Suzuki2013b,Suzuki2013a}.
Hence, when the system temperature is low and the spins seldom flip,
the simulation performs efficiently,
in a manner similar to continuous-time Monte Carlo algorithms.
However, it is generally less efficient on ordinary computers
compared with the ordinary heat-bath algorithm~\cite{Suzuki2013b}.

\section{Convergence}

In this section,
we numerically verify that the billiard dynamics yields
samples that converge to the true distribution
of the Ising model on a small lattice,
for which probability distributions can be computed.

\begin{figure}
\centering
\includegraphics{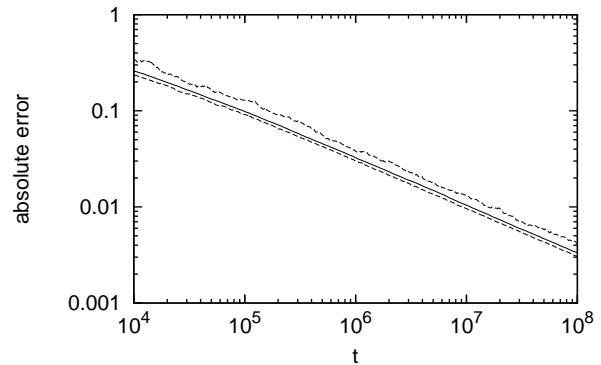}
\caption{The absolute errors of the empirical distribution
for the Ising model on a two-dimensional lattice of size $L=4$
with the periodic boundary condition at temperature $T=2.4$.
The absolute error from the true distribution at time $t$ is calculated for
the empirical distribution obtained from a trajectory until time $t$.
The solid line indicates the average absolute error
for 96 different trajectories,
and the dashed lines indicate the minimum and maximum absolute errors
among the trajectories.
All the lines decrease at a gradient of nearly $-1/2$.}
\label{fig:convergence}
\end{figure}

\begin{figure*}
\centering
\includegraphics{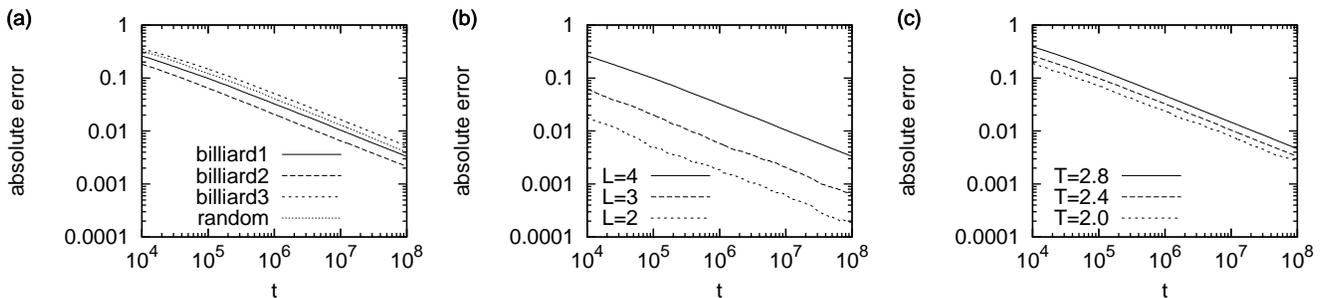}
\caption{The absolute errors of the empirical distribution obtained
by the billiard dynamics for the Ising model on a two-dimensional lattice
with the periodic boundary condition.
(a) The average absolute errors for different sampling algorithms.
The lines indicated by ``billiard1,'' ``billiard2,'' ``billiard3,'' and ``random''
correspond to Eqs.~(\ref{eq:dx}), (\ref{eq:dxps-}), (\ref{eq:dxp-s}),
and random Monte Carlo sampling, respectively.
(b) The average absolute errors
for different lattice sizes $L=2,3,4$ at $T=2.4$.
(c) The average absolute errors
for different temperature values $T=2.0,2.4,2.8$ with $L=4$.
All the lines shows the average absolute error for 96 different trajectories.
Overall, the errors of billiard sampling dynamics decay
very similarly to those of random sampling.}
\label{fig:tempsize}
\end{figure*}

Figure \ref{fig:convergence} shows the absolute error between
the empirical distribution observed from the billiard system
and the true distribution of the Ising model
on a two-dimensional lattice of size $L=4$.
The absolute error is defined as
\begin{equation}
\sum_{\bm{\sigma}}\left|P[\bm{\sigma}] - \frac{r_{\bm{\sigma}}(t)}{t} \right|,
\end{equation}
where $r_{\bm{\sigma}}(t)$ denotes the total time for which
the system takes on the state $\bm{\sigma}$ until time $t$.
The solid line in Fig.~\ref{fig:convergence} indicates
the absolute error averaged for 96 different initial values,
and the dashed lines indicate the minimum and maximum errors.
Note that the error is calculated for each orbit.
Therefore, Fig.~\ref{fig:convergence} shows that for all the 96 initial values
randomly chosen from the uniform distribution on $\{-1,+1\}^N\times[-1,+1]^N$,
the empirical distributions constantly converge to the true distribution.
All the lines have gradients nearly equal to $-1/2$,
thereby indicating that the convergence rate is almost $O(1/\sqrt{t})$.

Figure~\ref{fig:tempsize}(a) shows the absolute errors
for Eqs.~(\ref{eq:dx}), (\ref{eq:dxps-}), (\ref{eq:dxp-s}),
and random Monte Carlo sampling.
As for random Monte Carlo sampling, $N$ Monte Carlo steps are
regarded as one unit time.
The errors decrease in a similar manner in each case.
The constant biases between the algorithms are mainly due to
the difference in the switching frequencies of the algorithms.
This result numerically shows that
all the three definitions for $\mathrm{d}x/\mathrm{d}t$,
which are justified only intuitively in the previous section,
work consistently with Eq.~(\ref{eq:prob}).

Figure~\ref{fig:tempsize}(b) shows the absolute errors
for different lattice sizes $L=2$, $3$, and $4$.
The errors decrease at almost the same gradient.
For calculation for $L=2$ and $3$, we perturbed the speeds of
the units by multiplying constant values to Eq.~(\ref{eq:dx}).
Specifically, Eq.~(\ref{eq:dx}) for $i=1,\dots,N$ is multiplied
by $0.5+0.2i$ for $N=2\times 2$ and by $0.5+0.1i$ for $N=3\times 3$.
Without this perturbation, the errors sometimes stop decreasing,
which may be due to symmetries in these small systems composed of uniform units.
For example, in the case of $L=2$, two units in diagonal positions
receive exactly the same inputs from two other units
in the other diagonal positions.
Since it can be intuitively understood that the dynamics can be easily limited
to a subspace if all the units are uniform,
we perturbed the speeds and confirmed the convergence as expected.

Figure~\ref{fig:tempsize}(c) shows the average absolute errors
at different temperatures $T=2.0$, $2.4$, and $2.8$ for $L=4$.
The errors decrease with almost the same gradient.

To summarize, all the results show convergence to the true
distribution with the convergence rate almost in the order of $O(1/\sqrt{t})$.
Namely, the convergence rate is almost the same as that of
random Monte Carlo sampling, and slower than the order $O(1/t)$
of the herded Gibbs sampling \cite{Bornn2013}.

\section{Finite-size scaling analysis}

To examine the capabilities of the billiard dynamics for larger lattices, 
we apply it to finite-size scaling analysis (see, e.g., Ref.~\cite{Newman1999})
of the critical behavior of the Ising model.
In the context of CMLs, it is known that the universality class
depends on the updating rules \cite{Marcq1997}.
Therefore, it is intriguing to determine as to which universality class
the continuous-time billiard dynamics belongs to,
although it should belong to the Ising universality class
if the billiard dynamics truly yields samples from the Ising model precisely.

\begin{figure}
\centering
\includegraphics{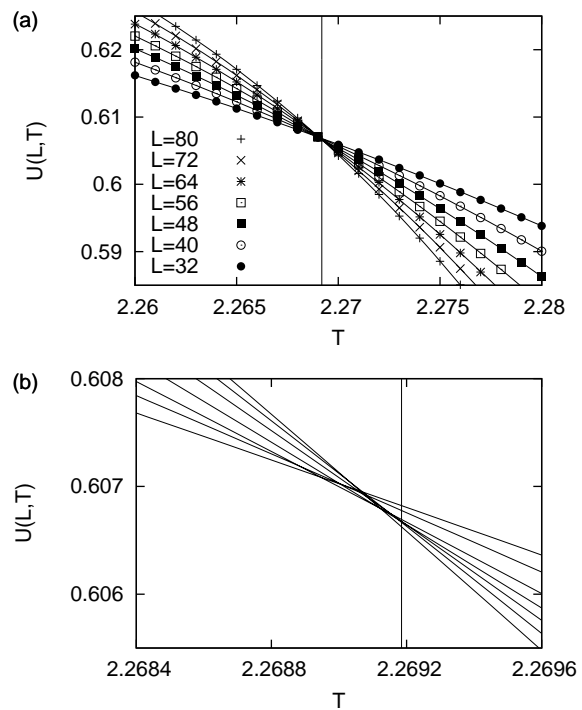}
\caption{The Binder cumulant as a function of the temperature $T$
around $T_\textrm{c}$ for $L=32$, $40$, $48$, $56$, $64$, $72$, and $80$.
(a) The points show the values of the Binder cumulant
calculated numerically by the billiard dynamics.
The statistics are calculated during $10^7$ unit time,
after initial $10^6$ unit time is skipped, for 96 different initial values.
The solid lines indicate least square fittings with cubic functions
for each lattice size.
(b) Magnification of the graphs in (a) around
the critical temperature $T_\textrm{c}$.
The intersections of the cubic fittings provides an estimate of
$T_\textrm{c}=2.2690\pm 0.0002$, which is consistent with the theoretical
critical temperature of $T_\textrm{c} = 2/\log(1+\sqrt{2})=2.2691\cdots$
of the Ising model on a two-dimensional lattice
indicated by the vertical line.
Note that the graphs for larger lattices tend to give better estimates.}
\label{fig:binder}
\end{figure}

\begin{figure*}
\centering
\includegraphics{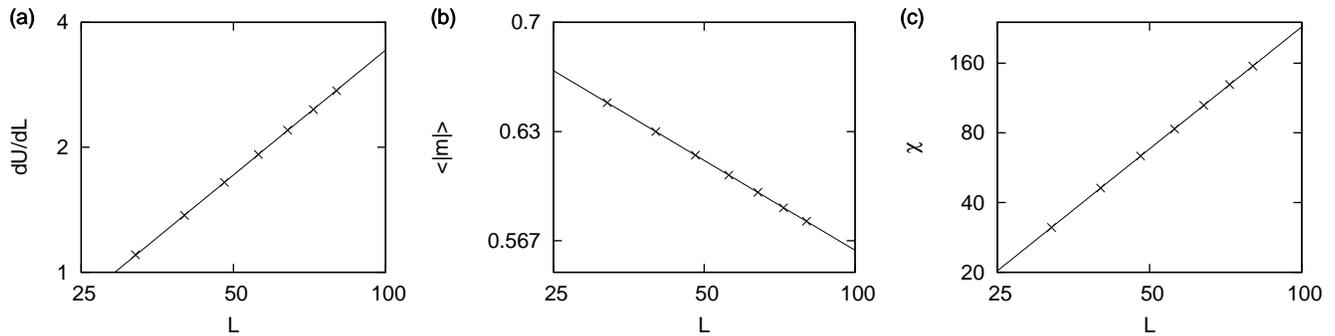}
\caption{Dependencies on lattice size $L$ of
(a) the gradient of the cubic fittings to the Binder cumulant
$\partial U/\partial T$,
(b) the mean absolute magnetization  $\langle|m|\rangle$,
and (c) the magnetic susceptibility $\chi$,
all calculated at the theoretical critical temperature $T=T_c$.
The crossings show the values obtained by numerical calculations.
These log--log plots give the estimates
(a) $1/\nu=0.993\pm 0.016$,
(b) $\beta/\nu=0.1248\pm 0.0004$,
and (c) $ \gamma/\nu=1.751\pm 0.004$.
These results are in agreement with the theoretical values of
the critical exponents $\nu=1$, $\beta=1/8$, and $\gamma=7/4$ of
the Ising model on a two-dimensional lattice.
The lines show least square fits by linear functions
with the theoretical gradient values.}
\label{fig:exponents}
\end{figure*}

We use the Ising model on a two-dimensional lattice of size $L=32,40,\dots,80$
with the periodic boundary condition.
To find the phase transition point,
we calculated the Binder cumulant \cite{Binder1981}
\begin{equation}
U(L,T) = 1 - \frac{\langle m^4 \rangle}{3 \langle m^2 \rangle^2},
\end{equation}
where $m$ denotes the magnetization $m=\sum_i\sigma_i/N$.
The graphs of the Binder cumulants for different lattice sizes $L$
are expected to intersect at the critical temperature $T_c$.
Figure \ref{fig:binder} shows the Binder cumulants calculated
for various lattice sizes around the theoretical critical temperature
and least square fittings with cubic functions to the calculated values.
The intersections of the polynomial fits give an estimate of $2.2690\pm 0.0002$,
which is consistent with the theoretical value of
$T_\textrm{c} = 2/\log(1+\sqrt{2})=2.2691\cdots$
of the Ising model on a two-dimensional lattice.

The exponent $\nu$ can be obtained from the derivative of
the Binder cumulant with respect to the temperature
at the critical temperature,
because according to the finite-size scaling theory,
the following relation holds:
\begin{equation}
\left.\frac{\partial U}{\partial T}\right|_{T=T_c}\propto L^{1/\nu}.
\end{equation}
Figure \ref{fig:exponents}(a) shows the derivatives of the polynomial fits
at the theoretical critical temperature.
The gradient of the log--log plot is estimated as $0.993\pm 0.016$,
which is consistent with
$\nu=1$ of the Ising model.
The exponents $\beta$ and $\gamma$ can be obtained similarly
from the absolute magnetization $\langle|m|\rangle$ and
magnetic susceptibility $\chi=N(\langle m^2\rangle - \langle|m|\rangle^2)$
at the critical temperature, using the following relation:
\begin{align}
\langle |m| \rangle &\propto L^{-\beta/\nu},\\
\chi &\propto L^{\gamma/\nu}.
\end{align}
From the log--log plots shown in Fig. \ref{fig:exponents}(b) and (c),
the critical exponents $\beta/\nu$ and $\gamma/\nu$ are estimated as
$0.1248 \pm 0.0004$ and $1.751 \pm 0.004$, respectively.
These values are consistent with
those of the Ising universality class: $\nu=1$, $\beta=1/8$, and $\gamma=7/4$.

\section{Potts model}

In this section, we extend the billiard dynamics to
spins that take more than two states,
and we apply it to the Potts model.
For simplicity, we describe the dynamics for the Potts model here;
it is straightforward to extend the dynamics to more general spin systems.

The Hamiltonian of the $q$-state Potts model is given by
\begin{equation}
E(\bm{\sigma})=-\sum_{\langle i,j\rangle}\delta(\sigma_i,\sigma_j),
\end{equation}
where $\sigma_i\in\{0,1,\dots,q-1\}$ denotes the state of the $i$th site
and $\delta(\cdot,\cdot)$ represents the Kronecker delta function.

We implement the Potts model
on the basis of the billiard sampling dynamics for the Ising model as follows.
We define an internal state $x_i\in \mathbb{R}/q\mathbb{Z} = [0,q)$
for each $i$th node,
which evolves according to the following differential equation:
\begin{equation}\label{eq:potts-dx}
\frac{\mathrm{d}x_i}{\mathrm{d}t}=
\exp\left(-\frac{1}{T}\sum_j\delta(\sigma_i,\sigma_j)\right).
\end{equation}
Note that the speed $\mathrm{d}x_i/\mathrm{d}t$ is always positive
and the two end points of $[0,q)$ are regarded as identical.
Therefore, $x_i$ continues to increase in the interval $[0,q)$,
and when it reaches $x_i=q$, it instantaneously jumps to $x_i=0$.
The state $\sigma_i$ of the $i$th node is determined from $x_i$ as
$\sigma_i = \lfloor x_i \rfloor$,
where $\lfloor x_i \rfloor$ denotes the largest integer
not greater than $x_i$.

Equation (\ref{eq:potts-dx}) is designed according to essentially
the same idea as Eq.~(\ref{eq:dx}) of the billiard dynamics of the Ising model.
Namely, the speed is determined so that it is proportional to
$P[\sigma_i\mid\bm{\sigma}_{\setminus i}]^{-1}$.
Therefore, when the probability is higher, the internal state moves more slowly,
and the system remains in such a state for a longer duration.

This system can be regarded as a coupled oscillator system,
because each internal state $x_i$ oscillates on a circle $\mathbb{R}/q\mathbb{Z}=[0,q)$
interacting with the neighboring nodes.
When $q=2$, this system is essentially equivalent
to the billiard dynamics for the Ising model.
However, for $q>2$, the internal state $(x_1,\dots,x_N)$
travels on a $N$-dimensional torus $[0,q)^N$,
and its dynamics cannot be reduced to a billiard system.
Note that, as mentioned in Ref.~\cite{Suzuki2013b},
another implementation by pseudo-billiard dynamics for the Potts model
is possible by using switched arrival systems~\cite{Chase1993}.

\begin{figure*}
\centering
\includegraphics{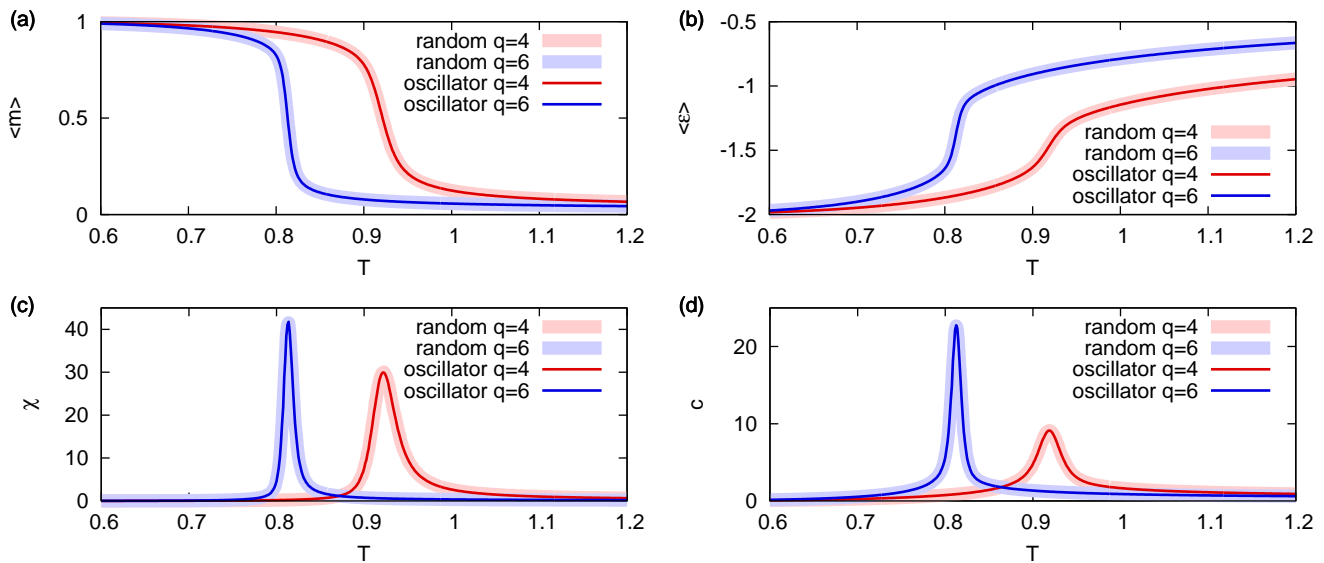}
\caption{Statistics of the Potts model on a two-dimensional lattice
of size $L=24$ with the periodic boundary condition
for $q=4$ (red lines) and $q=6$ (blue lines)
calculated by the heat-bath algorithm (thick lines)
and the oscillator sampling dynamics (thin lines).
(a) Average order parameter $\langle m\rangle$.
(b) Average energy per site $\langle\epsilon\rangle$.
(c) Magnetic susceptibility $\chi$.
(d) Specific heat $c$.
The statistics are calculated for $10^5$ unit time,
after initial $10^4$ unit time is skipped, for 100 different initial values.}
\label{fig:potts}
\end{figure*}

As a statistic that characterizes the behavior of the Potts model,
we define the order parameter as follows:
\begin{equation}
m = \max_s\frac{qN_s - N}{N(q-1)},
\end{equation}
where $N_s$ denotes the number of the sites taking on the state $s$.
The order parameter $m$ takes $1$ when all the spins take one
state ($\max_sN_s=N$) in the completely ordered phase,
and it takes $0$ when the spins take all the states equally ($N_s=N/q$)
in the completely disordered phase.

Figure \ref{fig:potts} shows statistics of the Potts model on a two-dimensional
lattice of size $L=24$ with the periodic boundary condition
for $q=4$ and $q=6$ calculated by the heat-bath algorithm
and the oscillator sampling dynamics.
The Potts model on a two-dimensional lattice is known to undergo
a second order phase transition if $q\le 4$
and a first order phase transition if $q>4$.
In Fig. \ref{fig:potts},
the average order parameter $\langle m\rangle$ and the average energy per site
$\langle\epsilon\rangle=\langle E \rangle/N$ as well as
their variances per site, the magnetic susceptibility
$\chi=N(\langle m^2\rangle - \langle m\rangle^2)$
and the specific heat
$c=N(\langle \epsilon^2\rangle - \langle\epsilon\rangle^2)$,
are shown.
In all the graphs, the lines for the two methods coincide with each other.

\section{Discussion}

\subsection{Extensions to continuous-valued spin models}

The implementation for the Potts model naturally leads us to
consider extensions to continuous-valued models such as the XY model.
Here, we discuss the possibilities of such extensions.

The XY model is composed of spins that take continuous-valued states
in $[0,2\pi)$.
The Hamiltonian of the XY model is given by
\begin{equation}
E(\bm{\theta})=-\sum_{\langle i,j\rangle}\cos(\theta_i,\theta_j),
\end{equation}
where $\theta_i\in[0,2\pi)$ denotes the state of the $i$th spin.
The XY model can be regarded as the limit $q\to\infty$
of the Potts model with the interaction term replaced by the cosine function.

One way to design sampling dynamics for the XY model
on the basis of the oscillator dynamics for the Potts model is as follows.
We define an internal state  $x_i\in \mathbb{R}/\mathbb{Z} = [0,1)$
for each $i$th node.
It evolves according to the following differential equation:
\begin{equation}\label{eq:xy-irrational}
\frac{\mathrm{d}x_i}{\mathrm{d}t}=
\exp\left(-\frac{1}{T}\sum_j\cos(\theta_i,\theta_j)\right).
\end{equation}
Since $\mathrm{d}x_i/\mathrm{d}t$ is always positive,
the internal state $x_i$ always increases from $0$ to $1$.
When it reaches $x_i=1$, it instantaneously jumps to $x_i=0$
and the state $\theta_i$ is updated as follows:
\begin{equation}
\theta_i \longleftarrow \theta_i + 2\pi\phi
\quad\text{and}\quad
x_i \longleftarrow 0
\qquad\text{when}\quad x_i=1,
\end{equation}
where $\phi$ is an arbitrary irrational number.
We use the golden mean  $\phi=(\sqrt{5}-1)/2$ in the numerical simulations.

\begin{figure*}
\centering
\includegraphics{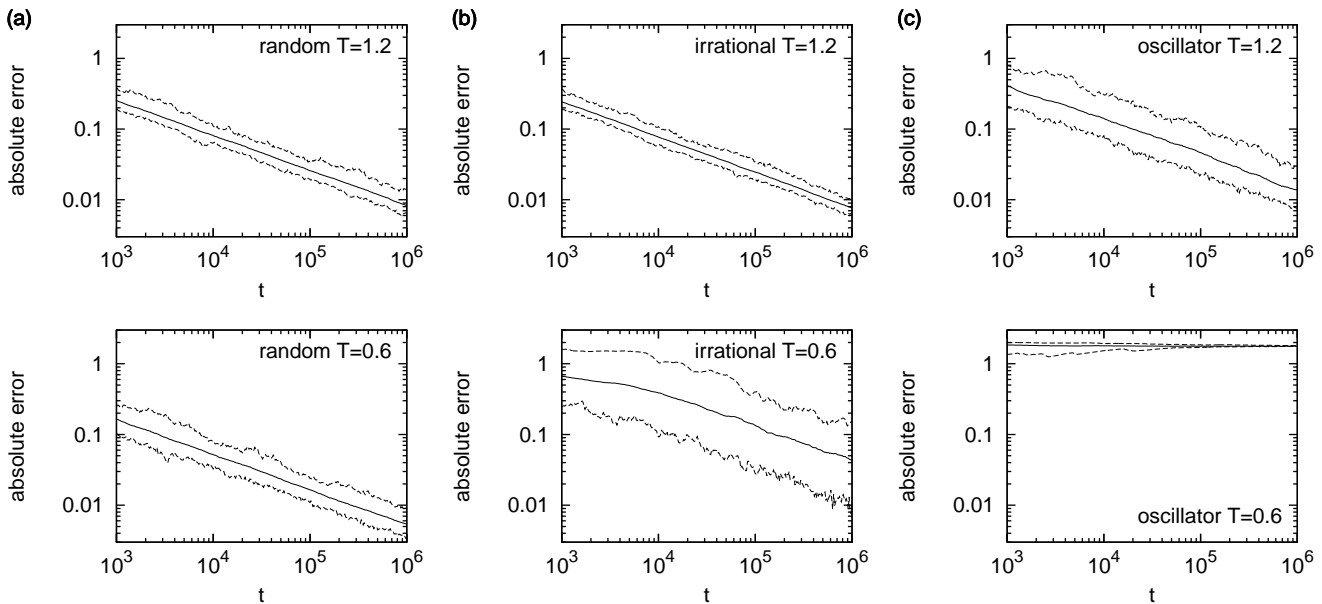}
\caption{The sampling dynamics for the XY model
on a two-dimensional lattice of $L=8$
at $T=1.2$ (top row) and $T=0.6$ (bottom row).
The absolute errors of empirical energy distributions
from the distribution obtained
by the Metropolis algorithm for $t=10^8$ are shown.
The empirical energy distributions are constructed from
samples until time $t$ generated by
the (a) Metropolis algorithm,
(b) irrational-rotation sampling dynamics, and
(c) coupled-oscillator sampling dynamics.
The solid lines indicate the average absolute error
for 96 different sample sequences,
and the dashed lines indicate the minimum and maximum absolute errors
among the sequences.}
\label{fig:xy-model}
\end{figure*}

As another possibility,
it is natural to consider the limit $q\to\infty$
in the implementation for the Potts model
(with the interaction term replaced by $\cos$).
Then, the internal state $x_i$ can be viewed
as the continuous-valued spin state $\theta_i$ of the XY model.
More specifically, the state $\theta_i$ evolves
according to the following differential equation:
\begin{equation}\label{eq:xy-oscillator}
\frac{\mathrm{d}\theta_i}{\mathrm{d}t}=
\exp\left(-\frac{1}{T}\sum_j\cos(\theta_i,\theta_j)\right).
\end{equation}
This equation defines a coupled oscillator system, although the interaction is
completely different from ordinary coupled oscillator systems
such as the Kuramoto model.
Since the state changes continuously,
the numerical simulation of this system is very different from
other systems we have considered.
In the following simulations, we integrate the differential equation
by the fourth-order Runge-Kutta method with
a time step of $0.01$.

We evaluate these methods by examining energy distributions
constructed from sample sequences for the XY model at $T=1.2$ and $0.6$,
where
energy values per site are discretized with a bin width of 0.01.
Since we do not have the true distribution,
we regard an empirical distribution constructed from samples until $t=10^8$
generated by the Metropolis algorithm as the ``true'' distribution,
where $N$ Metropolis steps are regarded as 1 unit time.
Figure \ref{fig:xy-model}(a), (b), and (c) shows
the absolute errors from the true distribution
of the empirical distributions obtained by the Metropolis algorithm,
the irrational-rotation sampling dynamics,
and the coupled-oscillator sampling dynamics, respectively.
For all the algorithms, sample sequences are obtained
by sampling uniformly at every 1 unit time.
As regards the irrational-rotation sampling (Fig.~\ref{fig:xy-model}(b)),
the error decreases almost
in the order of $O(\sqrt{t})$ even in the worst case.
While the coupled-oscillator sampling dynamics works well for $T=1.2$,
it does not work at all for $T=0.6$ (Fig.~\ref{fig:xy-model}(c)).
We have not succeeded in tracking down the cause of this result yet;
it is possible that Eq. (\ref{eq:xy-oscillator}) intrinsically
does not offer proper sampling dynamics
even if the equation can be integrated without numerical errors.
Although it is also possible that this may be only due to numerical errors,
this result at least shows that naive numerical integration does not work well.

To summarize, we have examined two types of sampling dynamics
designed for continuous-valued spin models.
While the irrational-rotation sampling dynamics (Eq.~(\ref{eq:xy-irrational}))
updates the state $\theta_i$ discretely,
the coupled-oscillator sampling dynamics (Eq.~(\ref{eq:xy-oscillator}))
updates the state $\theta_i$ continuously.
Although the irrational-rotation sampling seems promising,
further studies, especially on the differences between these two methods,
are necessary to understand
the capabilities and the limitations of the proposed dynamics.

\subsection{Spin echoes in the Ising model}

Thus far, we have mainly described the billiard and oscillator sampling dynamics
as a sampling method for classical spin systems.
However, we note that the dynamics itself is also interesting
as an abstract model for physical systems with
a large number of degrees of freedom
that can be related to both the CMLs and
the coupled oscillator systems.

\begin{figure}
\centering
\includegraphics{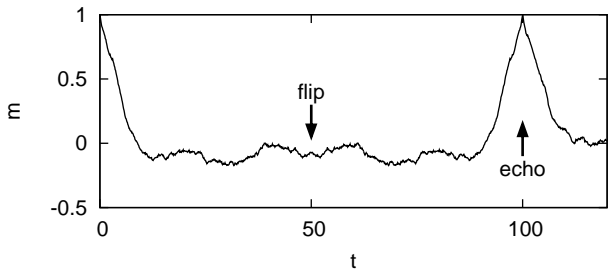}
\caption{Observation of a spin echo in the Ising model
on a two-dimensional lattice with size $L=64$ at $T=2.6$.
A time series of the magnetization $m$ sampled uniformly
at every $0.1$ unit time is shown.
At time $t=0$, all the spins are aligned.
At time $t=50$, all the internal states are flipped.
Subsequently, at time $t=100$, the spins become aligned again.}
\label{fig:spin-echo}
\end{figure}

One important aspect of the billiard sampling dynamics is
its reversibility,
while its macroscopic behavior as spin models is irreversible.
The microscopic origins of irreversible macroscopic behavior have been
an important topic in statistical physics since more than a century ago
(see, e.g., Ref.~\cite{Lebowitz1999}).
This topic has been discussed using models such as the Lorentz gas model
(or Sinai's billiard) \cite{Bunimovich1981},
multibaker maps \cite{Gaspard1992,Tasaki1995},
the Q2R automaton \cite{Stauffer2000},
and Nos\'e-Hoover thermostats \cite{Nose1984,Hoover1985,Hoover2012}.
The billiard sampling dynamics also exhibits both reversible microscopic behavior
and irreversible macroscopic behavior, and therefore, it can be a good example
for discussing this topic,
particularly if its chaotic dynamics is investigated in a more theoretical
manner in the future.
Note that the bakermap lattice \cite{Sakaguchi1999}
for the Ising model can be a good example as well,
because its dynamics is invertible and can be modified to be reversible.

As an example that demonstrates the reversibility,
we show in Fig.~\ref{fig:spin-echo} that
spin echoes \cite{Hahn1950} can be observed in the Ising model
with the billiard sampling dynamics.
We use the Ising model in the paramagnetic (disordered) phase.
At time $t=0$, all the spins are aligned to $+1$,
so that the magnetization is initially equal to $1$.
The initial internal state $\bm{x}$ is drawn
from the uniform distribution on $[-1,+1]^N$.
As the system equilibrates, the magnetization decreases
to values around zero.
At time $t=50$, we flip the signs of the internal states $\bm{x}$
instantaneously as $x_i \mapsto -x_i$,
while the states $\bm{\sigma}$ are kept unchanged.
After the flip, the simulation continues as if time is reversed.
The magnetization remains in the equilibrium state for a while;
however, it suddenly increases to $1$ at time $t=100$.
Namely, all the spins are aligned again at the moment,
contradictory to the second law of thermodynamics.
This result is a natural consequence of the reversible dynamics,
and theoretically, the memory of the initial state
can be restored after an arbitrarily long time.
However, it will practically become difficult to restore the initial state
after a long equilibration time on a large lattice,
due to the sensitive dependence on the initial conditions
of the chaotic dynamics.
This indicates that it will become almost impossible to
find a microscopic state that evolves contrary to the second law
of thermodynamics.

\section{Summary}

In this paper, we have numerically verified that
the billiard dynamics can generate samples from the Ising model
by examining the convergence and applying it to finite-size scaling analysis.
We also have extended the billiard dynamics to multi-valued and
continuous-valued spin models.
In all the simulations (except for the oscillator dynamics for the XY model),
the proposed dynamics works well
as a deterministic alternative to random Monte Carlo sampling.

It is considered important to examine more spin models
under various conditions on various network structures numerically.
However, because there are infinitely many possible models to examine,
we cannot clarify the capabilities and the limitations of this approach
only by the means of numerical simulations.
Therefore, we consider future studies on the theoretical foundations of
the sampling dynamics to be also important on the basis of the promising
results presented in this paper.

\begin{acknowledgments}
The author thanks Kazuyuki Aihara, Yoshihiko Horio, and Jun-ichi Imura
for their collaboration on earlier work on chaotic Boltzmann machines.
This study is supported by the Aihara Innovative Mathematical
Modelling Project, the Japan Society for the Promotion of Science
(JSPS) through the ``Funding Program for World-Leading Innovative R\&D
on Science and Technology (FIRST Program),'' initiated by the Council
for Science and Technology Policy (CSTP).
\end{acknowledgments}

\end{document}